\documentclass[journal]{new-aiaa}
\usepackage[utf8]{inputenc}
\usepackage{textcomp}
\usepackage{graphicx}
\usepackage{epstopdf}
\usepackage[version=4]{mhchem}
\usepackage{siunitx}
\usepackage{longtable,tabularx}
\usepackage[normalem]{ulem}
\usepackage{cancel}
\usepackage{bm, subfigure, footnpag, algorithm, algpseudocode, multirow, rotating, placeins, pdflscape}

\title{A Bayesian Benchmarking of GBEES \\ Applied to Outer Planet Orbiter Estimation}

\author{Benjamin L. Hanson\footnote{Corresponding Author; Graduate Student Researcher, Department of Mechanical and Aerospace Engineering; blhanson@uscd.edu.}}
\affil{University of California San Diego, La Jolla, California 92093}

\author{Todd A. Ely\footnote{Principal Navigation Engineer; todd.a.ely@jpl.nasa.gov.}}
\affil{Jet Propulsion Laboratory, California Institute of Technology, Pasadena, California 91109}

\author{Thomas R. Bewley\footnote{Professor, Department of Mechanical and Aerospace Engineering; tbewley@ucsd.edu.} and Aaron J. Rosengren\footnote{Assistant Professor, Department of Mechanical and Aerospace Engineering; arosengren@ucsd.edu.}}
\affil{University of California San Diego, La Jolla, California 92093}

\begin{document}

\footnotetext{Presented as Paper 25-194 at the AAS/AIAA Space Flight Mechanics Meeting, Kauai, Hi, Jan. 19-23, 2025 \cite{bH25}.}

\maketitle

\section{Introduction}
\lettrine{N}{avigation} has long adhered to an established assumption that is often underappreciated: spacecraft (S/C) state uncertainty remains Gaussian on the time scale of the measurement updates \cite{kalman}. This assumption, while nontrivial, is operationally validated via frequent telecommunication with the S/C, dividing the nonlinear trajectory into rectilinear segments. Adequate measurement cadences are assured for near-Earth S/C, but deep-space missions pose advanced challenges. To thoroughly explore outer planet systems, mission designers may leverage unstable, periodic orbits, the dynamics of which are often highly chaotic \cite{sR22,sC13}. Such chaoticity, dominated by third- and fourth-body perturbations, make linearizations accurate on much shorter time scales, thus requiring more frequent measurement updates and station-keeping maneuvers. However, in deep-space regimes, ground-based communication has significant round-trip light-time delays which may render this form of communication insufficient.

Historically, deep-space S/C have operated in stable, two-body orbits. However, proposed missions to outer planet systems, such as the Enceladus mission, suggest that the risk of operating an S/C in a dynamically unstable environment is outweighed by the scientific potential. The Enceladus S/C is expected to operate in an unstable, near-rectilinear halo orbit and perform a flyby within 10 km of the surface to perform \textit{in situ} analysis of the plumes of gas ejecting from the moon's surface \cite{ESA-Voyage-2050}. As such, it is likely that the instability of these conditions will cause the state uncertainty to become non-Gaussian faster than can be corrected by ground communication.

In this analysis, we aim to compare a fundamentally different estimation filter, Grid-based Bayesian Estimation Exploiting Sparsity (GBEES) \cite{tB12}, to a range of conventional and state-of-the-art filters. The application used for comparison is the propagation of the state uncertainty of a low-fidelity outer planet orbiter. This investigation is arranged into six sections. Section \ref{sec:sec_2} explains the filtering process and provides rationale for the filters compared in this study. Section \ref{sec:sec_3} presents a novel Bayesian approach to filter benchmarking, an alternative to the frequentist approach. Section \ref{sec:sec_4} defines the nonlinear dynamics and measurement models used in the Saturn-Enceladus DPO example. The numerical results are presented in Section \ref{sec:sec_5}. Finally, Section \ref{sec:sec_6} discusses in depth the results of the investigation and comments on the efficacy of GBEES as an onboard navigation filter.

\section{Overview of Sequential Single-update Filters}
\label{sec:sec_2}
The Sequential Single-update Filter (SSF) is the most widely used class of algorithms for S/C navigation and estimation. \textit{Sequential} refers to the way measurements are incorporated: one at a time as they become available, in contrast to batch filters, which process all data simultaneously. \textit{Single-update} denotes that each measurement is assimilated through a single application of Bayes' theorem\footnote{The particle flow filter \cite{flow}, a modern version of the Lagrangian approach, incrementally performs the Bayesian update by flowing the $M$ particles subject to a continuous parameter $\lambda$. Because this update step employs Bayes' rule multiple times per measurement, a methodological advantage that is unique to the particle flow filter, the authors believe its inclusion in this work would remove the one-to-one nature of the comparison.}. In this investigation, we label SSFs as either \textit{moment} filters or \textit{ensemble} filters, indicating the abstraction updated in the correction step (for clarification, moment filters in this context do not include those that propagate mixtures of moments or higher order moments).

The evolution of a stochastic process $\bm{X}(t)\in \mathbb{R}^d$ governed by a combination of deterministic and random effects can be described by the following stochastic differential equation:
\begin{equation}
\label{eq:SDE}
    d\bm{X}(t) = \bm{f}(\bm{X}(t), t)dt + \bm{q}(\bm{X}(t), t)d\bm{W}(t),
\end{equation}
where $\bm{f}(\bm{X}(t), t)$ is the drift vector, $\bm{q}(\bm{X}(t), t)$ is the diffusion vector, and $d\bm{W}(t)=\boldsymbol{\xi}(t)dt$ is a Wiener process, meaning $\boldsymbol{\xi}(t)$ is zero-mean, uncorrelated white noise. In continuous-time, the Fokker-Planck equation gives the evolution of the probability density function (PDF) $p(\bm{x}, t)$ of $\bm{X}(t)$ in Equation \eqref{eq:SDE} as follows:
\begin{gather}\label{eq:FPE}
    \frac{\partial p(\bm{x}, t)}{\partial t} = -\sum\limits_{i=1}^d\frac{\partial f_i(\bm{x},t)p(\bm{x},t)}{\partial x_i} + \frac{1}{2}\sum\limits_{i=1}^d\sum\limits_{j=1}^d\frac{\partial^2 Q_{ij}(\bm{x}, t)\,p(\bm{x},t)}{\partial x_i \partial x_j}
\end{gather} 
where $\bm{x} = (x_1, \dots, x_d)$ is a realization of $\bm{X}(t)$, $f_i(\bm{x}, t)$ is the $i^{\text{th}}$ component of the drift vector at realization $\bm{x}$ and time $t$, and $Q_{ij}(\bm{x}, t)$ is the $(i, j)^{\text{th}}$ element of the stochastic state disturbance matrix $Q = \bm{q}(\bm{x}, t)\bm{q}(\bm{x}, t)^T$. In general, Equation \eqref{eq:FPE} is non-integrable, and $p(\bm{x},t)$ cannot be described by a finite number of parameters.

At measurement epochs $t=t^{(k)}$, the PDF is updated via Bayes' theorem:
 \begin{gather}\label{eq:Bayes}
        p(\bm{x},t^{(k+)}) = \frac{p(\bm{y}^{(k)}|\bm{x})\,p(\bm{x},t^{(k-)})}{C},
\end{gather}
where $p(\bm{x},t^{(k+)})$ is the posterior, $p(\bm{y}^{(k)}|\bm{x})$ is the measurement likelihood, $p(\bm{x},t^{(k-)})$ is the prior, and $C$ is a normalization constant. Again, in general, the number of parameters necessary to represent $p(\bm{x},t^{(k+)})$ is not finite. Thus, the goal of the SSF is to approximate and propagate the full PDF with as few parameters as possible while incorporating information from measurements updates. 

We now list the filters GBEES is compared with in this analysis, as well as the rational for their inclusion. A large smooth bootstrap is used as a benchmark for both accuracy and efficiency. The Unscented Kalman Filter (UKF) \cite{UKF} is compared as a moment filter control, as its nonlinear prediction step makes it more accurate than other moment filters. The ensemble filters
compared are the Bootstrap Particle Filter (BPF) \cite{BPF} and the Ensemble Gaussian Mixture Filter (EnGMF) \cite{EnGMF}. The BPF is widely used, while the EnGMF and GBEES are more state-of-the-art. Recent efforts have focused on reducing the computational cost of determining the optimal bandwidth parameter for the EnGMF samples \cite{sY22, aP24}. Similarly, GBEES has recently been significantly computationally optimized \cite{bH24};  one objective of this investigation is determining if GBEES may be feasibly used as an onboard S/C navigation algorithm.

\section{Filter Benchmarking Criteria}
\label{sec:sec_3}

In this investigation, the performances of the selected SSFs are evaluated based on three criteria: accuracy is assessed on probability distribution similarity, degeneracy is assessed on an approximation of the effective sample size, and efficiency is assessed on computation time. Each SSF is compared to a truth implementation, discussed in Section \ref{sec:truth}.

\subsection{Accuracy metric}
The frequentist approach to filter evaluation assumes that many Monte Carlo (MC) runs of the filter may be performed to calculate an average accuracy and consistency. Practically, this may not be feasible due to time constraints. Instead, we perform a deterministic, Bayesian comparison of the filter-estimated distributions with the representative truth model. The Bhattacharyya coefficient ($\text{BC}$) \cite{BC}, a metric of the similarity between two probability distributions $P$ and $Q$, is defined as
\begin{gather}\label{eq:BC}
    \text{BC}(P,Q\,|\, \chi) = \sum_{\bm{x}\in \boldsymbol{\chi}} \sqrt{P(\bm{x}) Q(\bm{x})} \quad \text{where} \quad \chi \subseteq \mathbb{R}^d, \quad |\chi| < \infty, \quad \text{and} \quad 0\leq \text{BC} \leq 1,
\end{gather}
where $\chi$ is the domain and $|\chi|$ is the cardinality. $\text{BC}(P,Q\,|\, \chi)=0$ indicates complete dissimilarity while $\text{BC}(P,Q\,|\, \chi)=1$ indicates identicality. For this analysis, $P$ is the truth distribution and $Q$ is the filter-estimated distribution, with $\chi=\{\bm{x}_i\}^M_{i=1}$, where $\bm{x}_i$ are the members in the truth distribution and $M$ is the sample size. Nearest-neighbor interpolation is utilized for calculating $Q(\bm{x})$ when $\bm{x} \notin \text{support}(Q)$. If the compared filter is a moment filter, then a normalized distribution $Q$ is created by evaluating all points in $\chi$ on probability density $q(\bm{x})$.

\subsection{Degeneracy metric}
Particle degeneracy is a key drawback to many MC-based methods. A metric for measuring degeneracy commonly used in the literature is effective sample size (ESS) \cite{aD01, cR10}, approximated as
\begin{gather}
\label{eq:ESS}
    \widehat{\text{ESS}} = \frac{1}{\sum_{i=1}^M P_i^2} \quad \text{where} \quad 1 \leq \widehat{\text{ESS}}\leq M.
\end{gather}
where $P_i$ is the weight of ensemble member $\bm{x}_i$. For a uniformly weighted sample,
$\widehat{\text{ESS}}=M$, and for a completely degenerated sample where all weights but one are zero, $\widehat{\text{ESS}}=1$. Standard Lagrangian methods are generally either (a) degenerate over long propagation windows or (b) inefficient. In case (a), too few ensemble members are initialized at the start of the propagation period, and measurement updates cause probability to focus on a few particles in the ensemble. In case (b), degeneracy is planned for by initializing too many ensemble members, with the expectation that the number of probable particles will shrink to an acceptable amount after the measurement update. This results in inefficiencies at the initial stages of the propagation period. Using $\widehat{\text{ESS}}$, we can measure degeneracy in each of the ensemble filters.

\subsection{Truth implementation}
\label{sec:truth}
Large MC simulations are often used to represent point-mass approximations of the true uncertainty. There is a glaring problem with this: the standard MC does not allow for measurement updates, and resampling directly from the measurement likelihood results in loss of the prior. To address this issue, we represent the truth distribution as a large smooth bootstrap \cite{bootstrap, sW95}. To begin, ensemble members $\bm{x}^{(k-1)}_{i}$ are sampled from the initial Gaussian prior uncertainty and probability $P^{(k-1)}_{i}$ is calculated:
\begin{subequations}
\begin{gather}
    \bm{x}^{(k-1)}_{i} \sim \mathcal{N}\left(\bm{x}\,|\,\boldsymbol{\mu}^{(k-1)}; \boldsymbol{\Sigma}^{(k-1)}\right) \quad \text{and} \quad P^{(k-1)}_{i} = \frac{\mathcal{N}\left(\bm{x}^{(k-1)}_{i}\,|\,\boldsymbol{\mu}^{(k-1)}; \boldsymbol{\Sigma}^{(k-1)}\right)}{\sum_{j=1}^M\mathcal{N}\left(\bm{x}^{(k-1)}_{j}\,|\,\boldsymbol{\mu}^{(k-1)}; \boldsymbol{\Sigma}^{(k-1)}\right)}, \\
    \text{where} \quad \mathcal{N}\left(\bm{x}^{(k)}_{i}\,|\,\boldsymbol{\mu}^{(k)}; \boldsymbol{\Sigma}^{(k)}\right)=\frac{1}{\sqrt{(2\pi)^d\left|\boldsymbol{\Sigma}^{(k)}\right|}}\exp{\left(-\frac{1}{2}\left(\bm{x}_i^{(k)}-\boldsymbol{\mu}^{(k)}\right)^T{\boldsymbol{\Sigma}^{(k)}}^{-1}\left(\bm{x}_i^{(k)}-\boldsymbol{\mu}^{(k)}\right)\right)},
\end{gather}
\end{subequations}
where $|\cdot|$ is the determinant. By attributing probability to the initial ensemble members, more accurate comparisons between filter-estimated and truth distributions may be made. Once initialized, the ensemble members are then propagated via the system dynamics up to a measurement update epoch $t^{(k)}$. To assimilate information from the propagated prior with the measurement update while avoiding particle degeneracy, we perform a resampling using the GBEES posterior $\tilde{p}(\bm{x},t^{(k+)})$. Each grid cell in $\tilde{p}\left(\bm{x},t^{(k+)}\right)$ is treated as a Gaussian kernel with covariance $\boldsymbol{\Sigma}_*^{(k)}$ determined by Silverman's rule of thumb \cite{silverman}
\begin{gather}
\label{eq:silverman}
    \boldsymbol{\Sigma}_*^{(k)} = \left(\frac{4}{d + 2}\right)^{\frac{2}{d+4}}{M_*^{(k)}}^{-\frac{2}{d + 4}}\,\boldsymbol{\Sigma}^{(k)},
\end{gather}
where $M_*^{(k)}$ is the number of grid cells in $\tilde{p}\left(\bm{x},t^{(k+)}\right)$ and $\boldsymbol{\Sigma}^{(k)}$ is the likelihood covariance at $t=t^{(k)}$. New states are sampled from the GBEES posterior, then zero-mean Gaussian noise is superimposed with covariance from Eq. \ref{eq:silverman} and probability is calculated based on the likelihood moments at $t=t^{(k)}$:
\begin{gather}
    \bm{x}_i^{(k+)}\sim \tilde{p}\left(\bm{x}, t^{(k+)}\right) + \mathcal{N}\left(\bm{x}\,|\,\boldsymbol{0}; \boldsymbol{\Sigma}_*^{(k)}\right)\quad \text{and} \quad P^{(k+)}_{i} = \frac{\mathcal{N}\left(\bm{x}^{(k+)}_{i}\,|\,\boldsymbol{\mu}^{(k)}; \boldsymbol{\Sigma}^{(k)}\right)}{\sum_{j=1}^M\mathcal{N}\left(\bm{x}^{(k+)}_{j}\,|\,\boldsymbol{\mu}^{(k)}; \boldsymbol{\Sigma}^{(k)}\right)}.
\end{gather}
From here, the new samples are propagated until the next measurement update epoch. We highlight that, while sampling from the estimated distribution may introduce some bias towards GBEES, the measurements in this investigation have small uncertainties relative to the prior, making the GBEES posterior distributions nearly identical to the likelihood distributions.

\section{System Models}
\label{sec:sec_4}
\subsection{Dynamics model}
The contemporary limitations of GBEES render examples beyond 4D computationally uncompetitive compared with other SSFs. Thus, the authors sought outer planet orbiter trajectories that could reasonably be approximated by a low-fidelity model within the algorithm’s capabilities. One such family of trajectories is the Distant Prograde Orbit (DPO). DPOs are planar periodic orbits centered on the secondary body $M_2$ of the Circular Restricted Three-Body Problem (CR3BP), with stable and unstable invariant manifolds that form heteroclinic connections between $L_1$ and $L_2$ Lyapunov orbits, meaning a substantial volume of the three-body system may be explored by forming chains of unstable periodic orbits connected by low-energy transfers \cite{DPO, gM10, mG21}. Because of their planar nature, DPO trajectories may be represented by the Planar CR3BP (PCR3BP). The state and orbital dynamics of a S/C in the PCR3BP in the synodic frame are 
\begin{gather}
\label{eq:PCR3BP}
    \bm{x} = \begin{bmatrix}
        x \\ y \\ \dot{x} \\ \dot{y}
    \end{bmatrix}\quad \text{and} \quad \dot{\bm{x}} = \bm{f}(\bm{x}) = \begin{bmatrix}
        \dot{x} \\ \dot{y} \\ 2\dot{y} + x - \frac{(1-\mu)(x+\mu)}{r_1} -\frac{\mu(x-1+\mu)}{r_2} \\ -2\dot{x} + y - \frac{(1-\mu)y}{r_1} -\frac{\mu y}{r_2}
    \end{bmatrix},
\end{gather}
where $\mu = m_2/(m_1 + m_2)$ is the mass ratio, $m_i$ represents the mass of body $M_i$, and $r_i$ is the distance to body $M_i$. 

\subsection{Measurement model}
\label{sec:meas_model} 
Because GBEES is a finite volume method, the Courant–Friedrichs–Lewy (CFL) condition \cite{CFL} must be satisfied to ensure stability. The smaller the covariance of the initial measurement, the smaller the grid width must be to accurately represent the prior, and the smaller the time step must be to satisfy the CFL condition. Therefore, the efficiency of the algorithm is dependent on the magnitude of the initial covariance. For this reason, the authors sought a realistic measurement model for an outer planet orbiter with error larger than what is attainable through ground communication. However, ground communication may not be feasible for outer planet orbiters.

The operation of outer planet orbiters in the chaotic regimes of space proposed will require frequent measurement updates, possibly more frequent than can be provided by ground stations on Earth due to the significant time delays at outer planet distances. Instead, onboard Line-of-Sight (LoS) optical navigation may be used. LoS is based on the optical observations of visible celestial bodies and stars, obtained by the S/C via onboard cameras or star-trackers. By combining multiple consecutive images of a visible celestial body, or measurements from a camera as well as a star-tracker, certain LoS techniques can reconstruct range, azimuth, and range-rate, all relative to the celestial body \cite{sC23, jC15, iN21}. For this reason, we use a low-fidelity model of a Line-of-Sight (LoS) optical navigation technique, with measurement function
\begin{gather}
    \bm{y} = \begin{bmatrix}
        \rho \vspace{0.1 cm}\\ \theta \\ \dot{\rho}
    \end{bmatrix} = \bm{h}(\bm{x}) = \begin{bmatrix}
        \sqrt{(x-1+\mu)^2+y^2} \vspace{0.1 cm}\\ \tan^{-1}\left(\frac{y}{x-1+\mu}\right) \vspace{0.1 cm}\\ \frac{(x-1+\mu)\dot{x} + y\dot{y}}{\rho}
    \end{bmatrix}, \quad R = \begin{bmatrix}
        \sigma_{\rho}^2 & 0 & 0\\
        0 & \sigma_{\theta}^2 & 0 \\
        0 & 0 & \sigma_{\dot{\rho}}^2
    \end{bmatrix},\quad 
    \begin{cases}
        \sigma_{\rho} = 20 \text{ km} \\
        \sigma_{\theta} = 1.74533\text{e-}2 \text{ rad} \\ 
        \sigma_{\dot{\rho}} = 2.0\text{e-}3 \text{ km/s}
    \end{cases}
    ,
\end{gather}
where $\rho$ is the range, $\theta$ is the azimuth angle, and $\dot{\rho}$ is the range-rate (all relative to $M_2$). This measurement error is within the realm of the magnitudes found by Casini et al. \cite{sC23} and Bella et al. \cite{sB22} in their deep space LoS analyses.

\section{Numerical Results}
\label{sec:sec_5}
\begin{figure}[t]
	\centering\includegraphics[width=0.9\textwidth, trim=1cm 0cm 1cm 0cm]{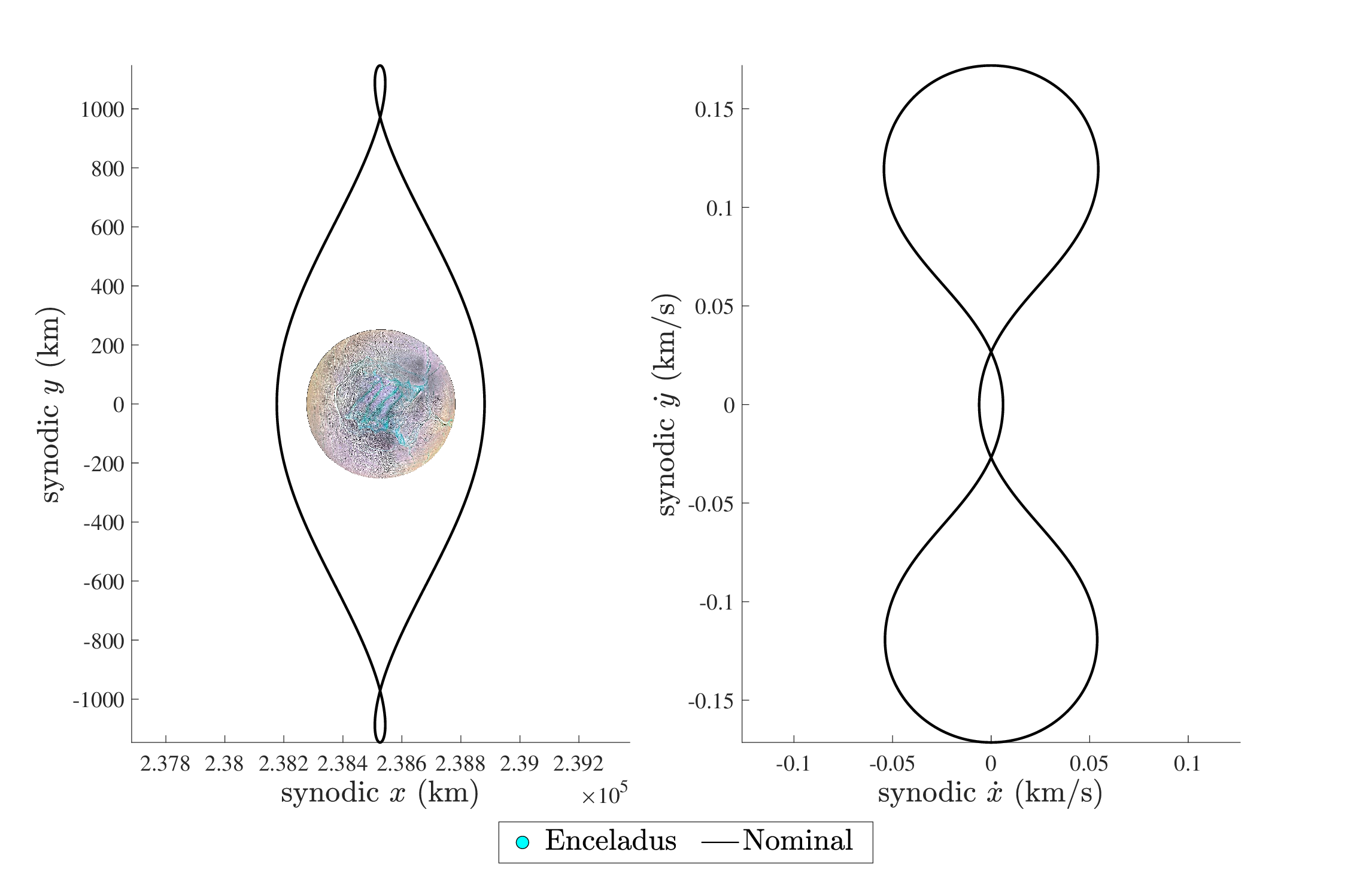}
	\caption{A Saturn-Enceladus DPO in the synodic frame; (\textit{left}) position and (\textit{right}) velocity.}
	\label{fig:SaEn_DPO_synodic_trajectory}
\end{figure}
\subsection{Trajectory properties}
The selected SSFs are evaluated on a Saturn-Enceladus Distant Prograde Orbit (DPO) with initial state and covariance
\begin{gather}
\label{eq:DPO_IC}
    \bm{x}^{(0)} = \begin{bmatrix}
        1.001471 \\ -1.751810\text{e-}5\\
        7.198783\text{e-}5 \\
        1.363392\text{e-}2
    \end{bmatrix} = \begin{bmatrix}
        238879.876159 & \text{(km)} \\ -4.178575 & \text{(km)} \\
        9.079038\text{e-}4 & \text{(km/s)} \\
        1.719497\text{e-}2 & \text{(km/s)}
    \end{bmatrix}, \quad
    \Sigma^{(0)} = \begin{bmatrix}
        \sigma_{\rho}^2 & 0 & 0 & 0 \\ 
        0 & \sigma_{\rho}^2 & 0 & 0 \\ 
        0 & 0 & \sigma_{\dot{\rho}}^2 & 0 \\ 
        0 & 0 & 0 & \sigma_{\dot{\rho}}^2
    \end{bmatrix}.
\end{gather}
Propagating Equation \eqref{eq:DPO_IC} with the dynamics from Equation \eqref{eq:PCR3BP} results in the trajectory in Figure \ref{fig:SaEn_DPO_synodic_trajectory}. Other properties of the system and trajectory may be found in the JPL Three-Body Periodic Orbit Catalog (trajectory ID: 1176) \cite{jpl_three_body_catalog}.

\subsection{Filter parameters}
The filter parameters are provided in Table \ref{tab:parameters}. For all filters, process noise is negligible, i.e., $Q=0$. The initial size $M^{(0)}$ of the BPF, EnGMF, GBEES are equal, while the size of the truth ensemble is much larger. The measurement cadence $\Delta t_{\bm{y}}$ evenly splits the propagation period into 4 segments. As a rule of thumb, the GBEES probability threshold $p_{\text{thresh}}$ is set such that $>75\%$ of the initial grid cells exceed $p_{\text{thresh}}$, and the grid width vector $\boldsymbol{\Delta}$ is set to half of the initial Cartesian standard deviation in all directions. The truth, UKF, BPF, and EnGMF are numerically integrated with Runge-Kutta 8(7) \cite{RK87}. 

\begin{table}[!ht]
\centering
\caption{Parameters of Filters}
\label{tab:parameters}
\resizebox{.99\hsize}{!}{$
\begin{tabular}{ c c c c }
\hline
\hline
\textbf{Model} & \multicolumn{1}{c}{\textbf{Propagation Scheme}}                                                                                                                                           & \multicolumn{1}{c}{$\bm{M^{(0)}}$} & \textbf{Misc.}                          \\ \hline
Truth    & \multicolumn{1}{c}{\multirow{5}{*}{\begin{tabular}[c]{@{}c@{}}RK$8(7)$\\ RK$8(7)$\\ RK$8(7)$\\ RK$8(7)$ \\ 2nd-order fully discretized \\ \end{tabular}}} & \multicolumn{1}{c}{5e+5}         &                                 \\ 
UKF   & \multicolumn{1}{c}{}                                                                                                                                                                          & \multicolumn{1}{c}{$2d + 1 = 9$}          & $\alpha=1\text{e-}3$, $\beta =2$, $\kappa =0$ \\
BPF   & \multicolumn{1}{c}{}                                                                                                                                                                          & \multicolumn{1}{c}{28561}        &                         \\ 
EnGMF & \multicolumn{1}{c}{}                                                                                                                                                                          & \multicolumn{1}{c}{28561}        &                      \\ 
GBEES & \multicolumn{1}{c}{}                                                                                                                                                                          & \multicolumn{1}{c}{28561}        & $p_{\text{thresh}} = 1\text{e-}7$, $\boldsymbol{\Delta} = [10 \text{ km}, 10 \text{ km}, 1\text{e-}3 \text{ km/s}, 1\text{e-}3 \text{ km/s}]$                      \\ \hline
\hline
\end{tabular}
$}
\end{table}

\begin{figure}[!t]
	\centering\includegraphics[width=0.9\textwidth, trim=1.5cm 0cm 1.5cm 0cm]{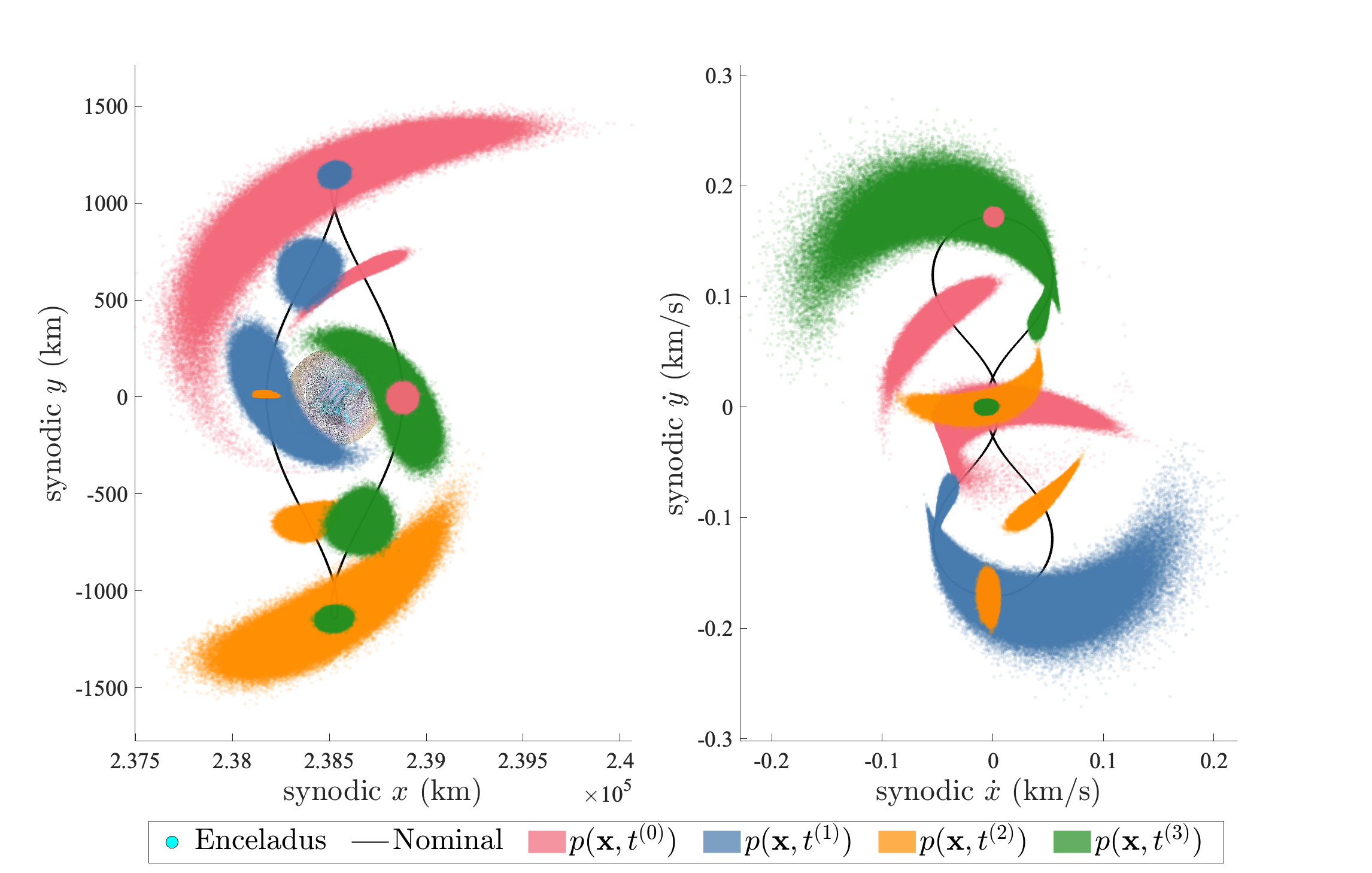}
	\caption{Saturn-Enceladus DPO true synodic state uncertainty.}
	\label{fig:SaEn_DPO_1176_mc}
\end{figure}

\subsection{Uncertainty propagation}
With the framework set, the state uncertainty of the Saturn-Enceladus DPO is propagated using the selected SSFs. Figure \ref{fig:SaEn_DPO_1176_mc} displays the true uncertainty distribution propagated by the large, smooth bootstrap. Both position and velocity curves become highly non-Gaussian at the particular measurement cadence, as expected. The color changes throughout indicate when a measurement update has occurred, where the posterior $p(\bm{x}, t^{(k+)})$ is the resultant assimilation of the prior $p(\bm{x}, t^{(k-)})$ and the likelihood distribution $p(\bm{y}^{(k)}|\bm{x})$. The distributions shown are temporally spaced to optimize visualization.

\begin{figure}
	\centering\includegraphics[width=0.99\textwidth, trim=0cm 0 0cm 0]{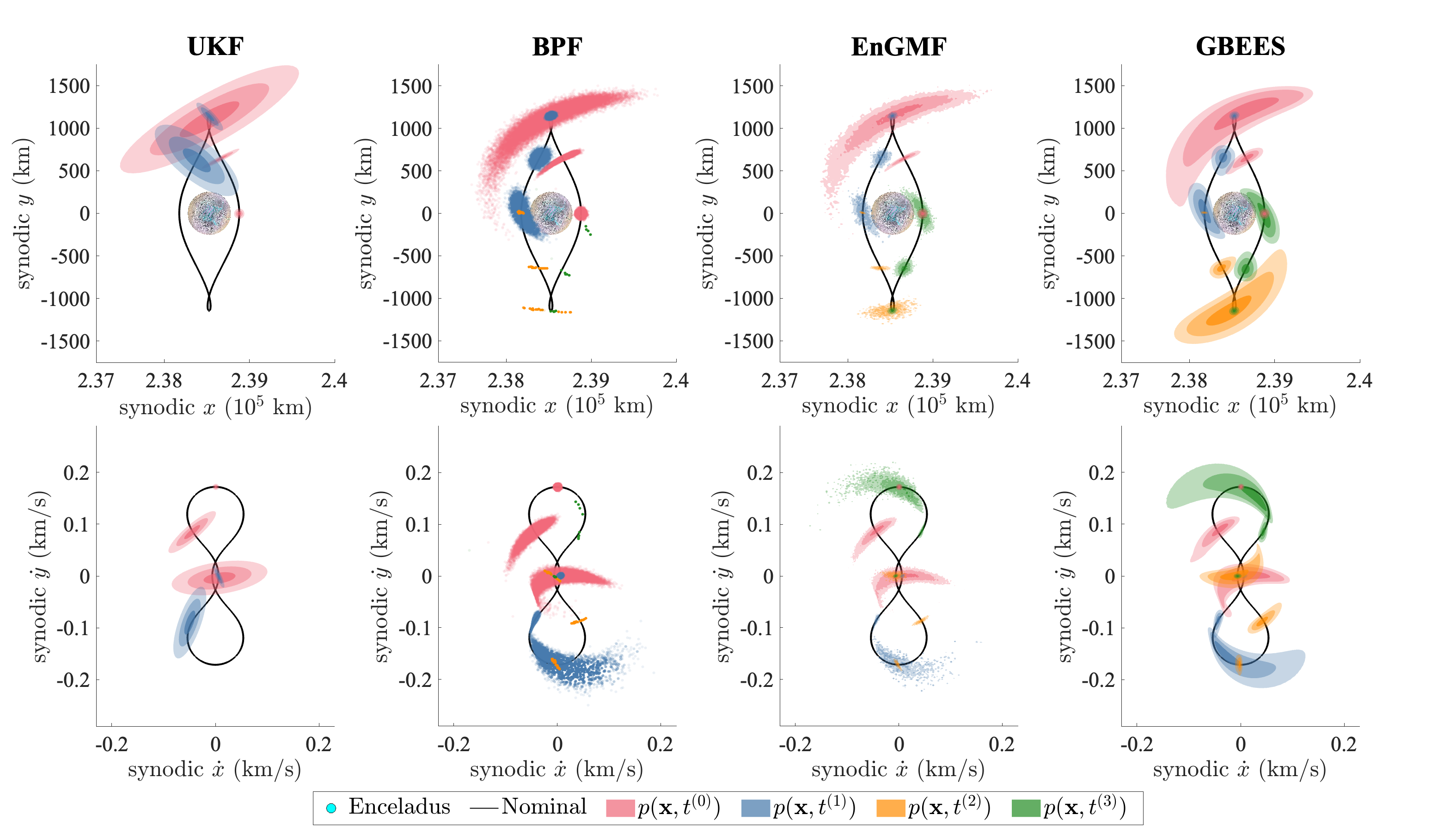}
	\caption{Saturn-Enceladus DPO synodic state uncertainty propagated by the UKF, BPF, EnGMF, and GBEES.}
	\label{fig:SaEn_DPO_1176_comp}
\end{figure}

The results of the selected SSFs are presented in Figure \ref{fig:SaEn_DPO_1176_comp}. The UKF implementation diverges in the prediction step after the first nonlinear measurement update, thus the absent PDFs beyond this epoch. The BPF implementation becomes highly degenerate after the three measurement updates, resulting in only 6 unique particles by the end of the propagation period. The EnGMF ensemble means, covariances, and weights are utilized to create isocontours by sampling points across the domain space and superimposing the probability from each component. The isocontour levels are set to $[0.68, 0.95, 0.997]$, \textit{a la} the empirical rule. The isocontours for GBEES are also set to the empirical rule values. We now aim to quantitatively compare these results using the defined metrics.

\begin{figure}[t]
    \centering
    \includegraphics[width=0.95\textwidth, trim=1.5cm 0 1.5cm 0]{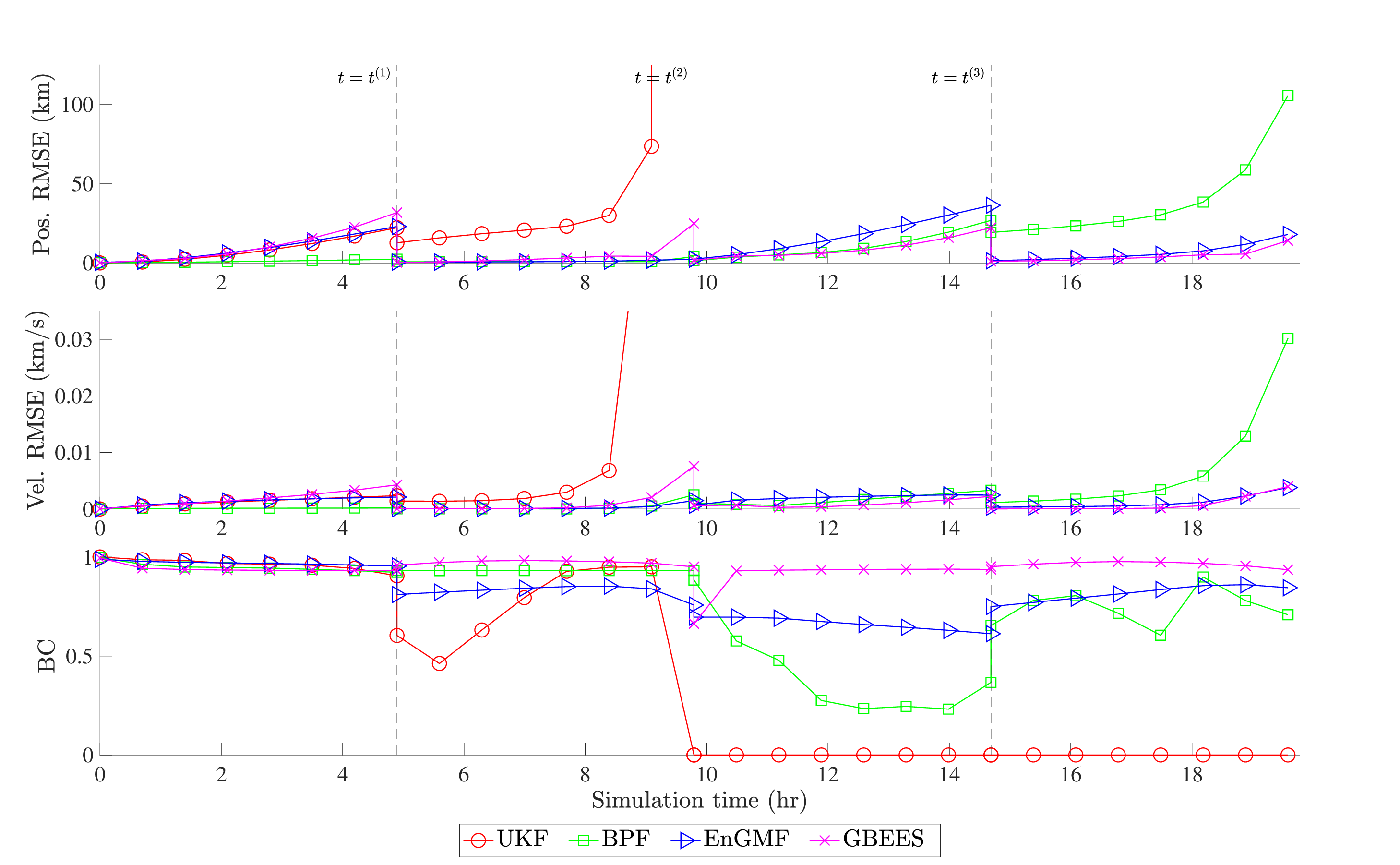}
    \caption{(\textit{top}) Position RMSE, (\textit{middle}) velocity RMSE, and (\textit{bottom}) $\text{BC}$ time histories for UKF, BPF, EnGMF, and GBEES implementations. Measurement update epochs are marked by vertical dotted lines.}
    \label{fig:SaEn_DPO_1176_metric}
\end{figure}

\begin{figure}[!t]
    \centering
    \includegraphics[width=0.9\textwidth, trim=2cm 0 2cm 0]{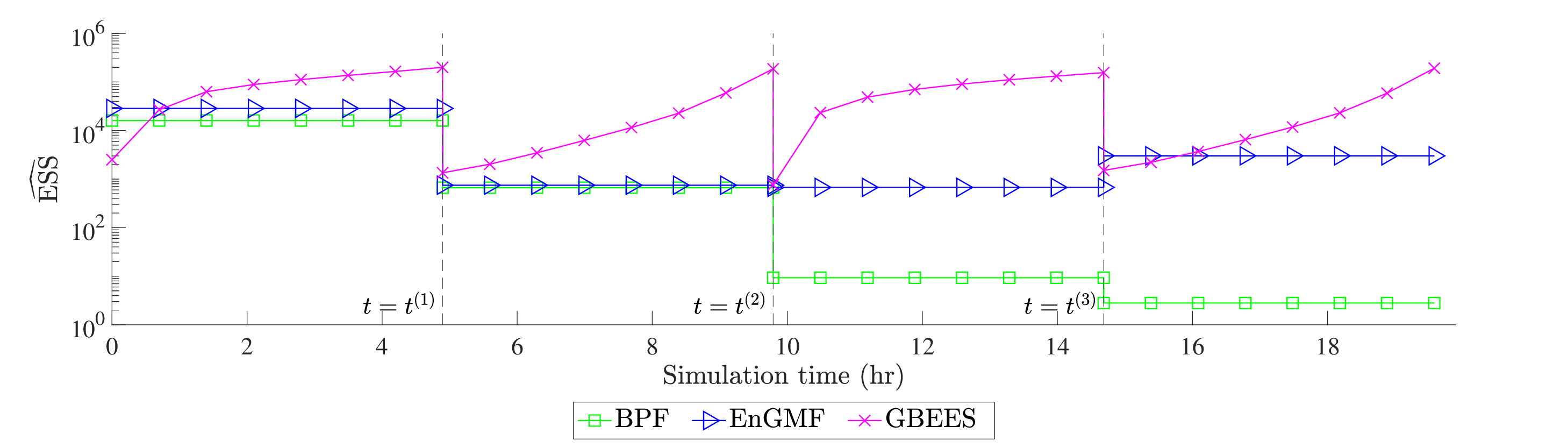}
    \caption{$\widehat{\text{ESS}}$ of the BPF, EnGMF, and GBEES implementations. Measurement update epochs are marked by vertical dotted lines.}
    \label{fig:SaEn_DPO_1176_ESS}
\end{figure}

Figure \ref{fig:SaEn_DPO_1176_metric} displays the time histories of the quantitative metrics of comparison for the UKF, BPF, EnGMF, and GBEES. Table \ref{tab:results} lists the final position root mean square error (RMSE), velocity RMSE, and $\text{BC}$ for each filter. The $\widehat{\text{ESS}}$ time histories of the ensemble SSFs are compared in Figure \ref{fig:SaEn_DPO_1176_ESS}. Due to the lack of process noise, for the BPF $M^{(k+1)} \leq M^{(k)}$, meaning each measurement update pushes the BPF towards degeneracy. Because the EnGMF resamples at each measurement epoch using the Gaussian sum filter update step, $M^{(k+1)} = M^{(k)}$. GBEES adaptively grows the discretized grid with the uncertainty, thus it avoids degeneracy throughout the prediction phase.

\begin{table}[!ht]
\centering
\caption{Final position RMSE, velocity RMSE, and $\text{BC}$ for selected SSFs}
\label{tab:results}
\begin{tabular}{ c c c c }
\hline
\hline
\textbf{Model} & \textbf{Position RMSE (km)} & \textbf{Velocity RMSE (km/s)} & \textbf{BC} \\ \hline
UKF   &  $1.835640\text{e+}6$                & $1.141540\text{e+}2$                      & 0.000   \\ 
BPF   &  $1.056313\text{e+}2$ & $3.016343\text{e-}2$ & 0.708 \\ 
EnGMF & $1.801216\text{e+}1$ & $3.789827\text{e-}3$ & 0.845 \\
GBEES & $1.414935\text{e+}1$ & $3.960708\text{e-}3$ & 0.937 \\ 
\hline
\hline
\end{tabular}
\end{table}

\begin{figure}[!t]
    \centering
    \includegraphics[width=0.7\textwidth, trim=3cm 0 3cm 0]{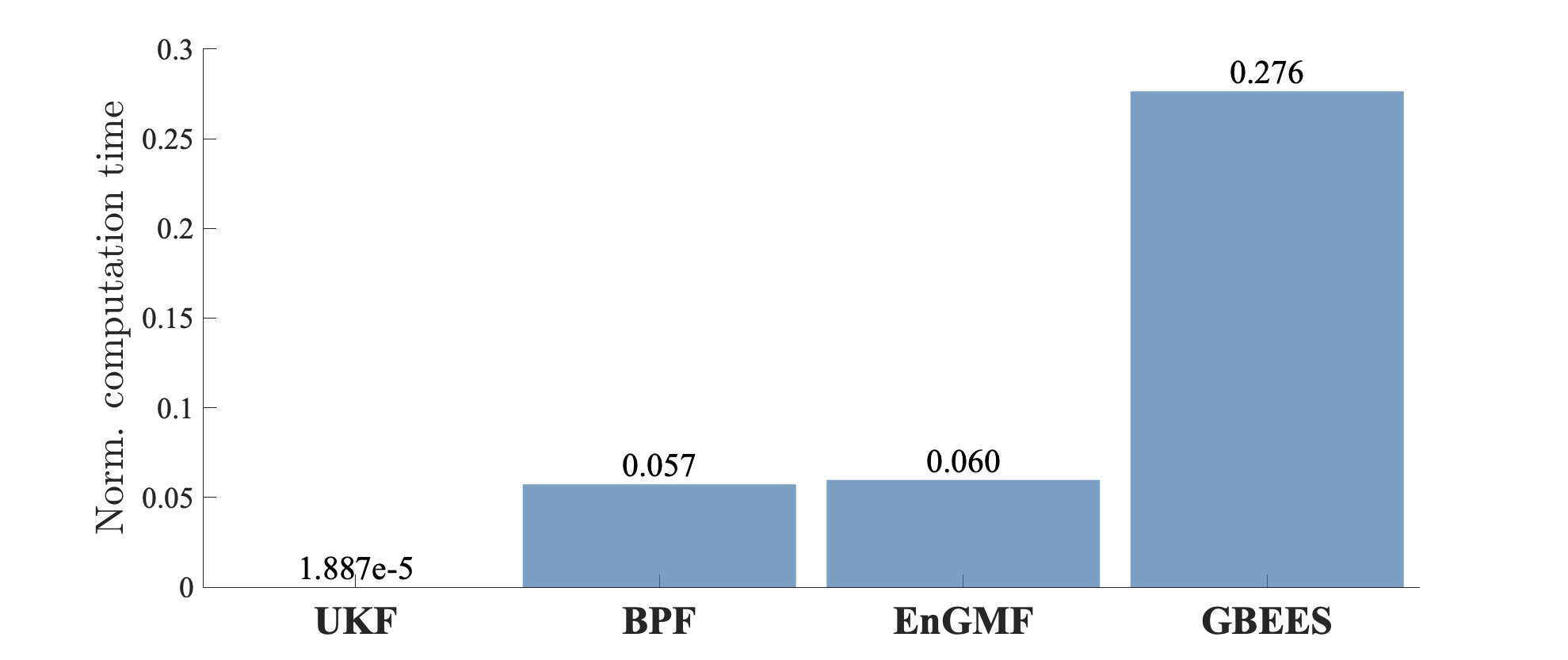}
    \caption{Normalized computation time of the UKF, BPF, EnGMF,  and GBEES implementations relative to the truth model.}
    \label{fig:SaEn_DPO_1176_runtime}
\end{figure}

The computational efficiency metric is normalized computation time, provided in Figure \ref{fig:SaEn_DPO_1176_runtime}. All models are run in C or C++ on a single core of a 3.49 GHz Apple M2 Max chip. The filter computation times are normalized by the truth model computation time, which takes $\sim$4.00 hours to perform the 19.58-hour propagation. As expected, the UKF implementation is the fastest. The BPF is about $5\times$ faster than GBEES, and the EnGMF is slightly slower than the BPF due to the intricate update step, but is much more accurate with the same size of ensemble.  

\section{Conclusion}
\label{sec:sec_6}
Conventional S/C navigation algorithms approximate uncertainty as the first and second central moments. This approximation has sufficed for all missions hitherto due to frequent measurement correction via telecommunication. However, future missions plan to operate outer planet orbiters in unstable regimes where the conventional may be insufficient due to fast-changing state uncertainty and round-trip light-time delays; in these cases, ensemble filters are necessary. Contemporary ensemble SSFs, while more accurate than their moment counterparts, can be inefficient when applied to high-dimensional systems and are prone to particle degeneracy. For practical onboard application, more robust solutions are required. 

In this investigation, the capabilities of a fundamentally different ensemble filter, Grid-based Bayesian Estimation Exploiting Sparsity, are compared to the UKF, BPF, and EnGMF via a Bayesian inference framework. The initially Gaussian state uncertainty of an unstable Saturn-Enceladus Distant Prograde Orbit is propagated for a full period, or about 19.58 hours, receiving nonlinear measurement updates every fourth of the period, or about 4.895 hours. The UKF implementation diverges shortly after the first measurement update. The BPF implementation estimates the uncertainty more accurately for longer than the UKF, but succumbs to particle degeneracy with each measurement update. The EnGMF resampling procedure disperses the probability more evenly across the distribution, making it an effective technique when process noise is low.  GBEES performs most consistently, returning a $\text{BC}$ value near 1 for the entirety of the propagation, finishing with position and velocity error nearly within 1$\sigma$ of the measurement uncertainty, and demonstrating degeneracy avoidance through the prediction phase, as indicated by the effective sample size.

The primary limitation of GBEES is its computational inefficiency. For a 4D problem, it is $5\times$ slower than the other ensemble filters evaluated in this study. As such, onboard S/C navigation via GBEES in its current form is impractical. However, the authors propose an alternative role for GBEES. This work highlights the challenges of using a MC method as a truth model when measurement updates are present. Instead, we suggest that GBEES serves as a high-fidelity ground-based truth model for validating faster filters, akin to how highly complex computational fluid dynamics models are used to simulate airflow in aircraft design. Although future efforts will still aim to improve the efficiency of GBEES, its primary utility may lie in this validation role.

\section*{Funding Sources}
This investigation was supported by the NASA Space Technology Graduate Research Opportunities Fellowship (80NSSC23K1219). The research was carried out in part at the Jet Propulsion Laboratory, California Institute of Technology, under a contract with the National Aeronautics and Space Administration (80NM0018D0004).

\section*{Acknowledgments}
The initial findings of this research were first presented at the 2025 AAS/AIAA Space Flight Mechanics Meeting in Kauai, HI. We gratefully acknowledge the valuable feedback and insights provided by attendees, which helped refine and enhance this work.

\bibliography{hanson_jgcd_rev2}
\end{document}